\documentclass[preprint,aps,amsmath,amssymb,nofootinbib,tightenlines,superscriptaddress]{revtex4}

\usepackage{graphicx}% Include figure files
\usepackage{axodraw}
\usepackage{dcolumn}% Align table columns on decimal point
\usepackage{bm}% bold math

\newcommand{\be}{\begin{equation}}
\newcommand{\ee}{\end{equation}}
\newcommand{\bea}{\begin{eqnarray}}
\newcommand{\eea}{\end{eqnarray}}

\newcommand{\tanb}{\ensuremath{\textrm{tan }  \beta} }
\newcommand{\fom}{\ensuremath{F/M_{mess}} }

\makeatletter       % Macros for arabic section numbering

\renewcommand{\p@subsection}{}

\renewcommand{\p@subsubsection}{}

\makeatother

\begin{document}

%\preprint{KEK-TH-XXXX}

\title{Dark Matter in Gauge Mediation
 from Emergent Supersymmetry}% Force line breaks with \\

\author{Siew-Phang Ng}%
\email{spng@physics.utoronto.ca}%
\affiliation{Department of Physics, University of Toronto,
Toronto, Ontario M5S 1A7, Canada.}

\author{Nobuchika Okada}%
\email{okadan@post.kek.ac.jp}%
\affiliation{Theory Division, KEK,
Oho 1-1, Tsukuba, Ibaraki, 305-0801, Japan.}
\affiliation{Department of Physics, University of Maryland,
College Park, Maryland 20742, USA.}

\date{\today}% It is always \today, today,
             %  but any date may be explicitly specified

\begin{abstract}
We investigated the viability of neutralino dark matter in the gauge mediation from emergent supersymmetry proposal. In this proposal, supersymmetry is broken at Planck scale and consequently, the gravitino is superheavy and completely decouples from the low energy theory. Squarks and sleptons obtain their soft masses dominantly through gauge mediation with other mechanisms highly suppressed. The lightest supersymmetric partner, in contrast to traditional gauge mediation, is a neutralino which is also a dark matter candidate. By explicit calculation of the low energy spectra, the parameter space was constrained using the WMAP observed relic density of dark matter, LEP2 Higgs mass bounds, collider bounds on supersymmetric partners and exotic B-meson decays. We found that the model has intriguing hybrid features such as a nearly gauge-mediated spectrum (the exception being the superheavy gravitino) but with a dominant  mSUGRA-like bino-stau coannihilation channel and at large $\tan \beta$, A-resonance-like annihilation.
\end{abstract}

%\pacs{Valid PACS appear here}% PACS, the Physics and Astronomy
                             % Classification Scheme.
%\keywords{Suggested keywords}%Use showkeys class option if keyword
                              %display desired
\maketitle

\section{Introduction}

Supersymmetry (SUSY) is one of the most promising way of extending the standard model to solve the gauge hierarchy problem\cite{SUSY}.  Since none of the superpartners have been observed in current experiments, SUSY should be broken but only so far that the superpartners remain unobservable while still maintaining the solution to the gauge hierarchy problem. However, soft SUSY breaking terms are also severely constrained to be almost flavor-blind and CP-invariant.  Thus, the SUSY breaking has to be mediated to the visible sector in some clever way so as not to induce too large CP and flavor violation effects. A variety of mechanisms to achieve such a viable mediation of SUSY breaking have been proposed\cite{Luty}.  With the impending commissioning of the Large Hadron Collider (LHC), it is more important than ever that the phenomenology of these various mechanisms be thoroughly explored so that we could identify the new physics that is expected to be found at the  TeV scale. Arguably, one of the most generic signatures of new physics is that of large missing transverse energy as a result of a very weakly interacting neutral stable particle exiting the detector . With standard model of cosmology needing a sizable amount of cold dark matter for structure formation and to populate galactic haloes, it is an extremely attractive and elegant proposition to explore the possibility that this weakly interacting neutral stable particle is one and the same as the constituent of dark matter. However, broad sweeping statements with multiple hidden assumptions have often been made about what this stable particle might tell us about the nature of the new physics. In this paper, we shall show the phenomenological viability of a minimal model of ``Gauge Mediation from Emergent Supersymmetry'' (GMES) that challenges conventional understanding of what it means to have gauge-mediated or minimal supergravity (mSUGRA) models.

In typical gravity-mediated scenarios, the gravitino is of the scale of the soft masses and because of its long lifetime, the late time gravitino decays lead to entropy production that would mess up big bang nucleosynthesis. Normal gauge mediation\cite{GiuRat}, on the other hand, gives much smaller gravitino masses and according to Ref.\cite{Viel}, would evade all astrophysical and cosmological bounds if it has a mass smaller than 16 eV. If one considers the possibility that there are additional cosmological moduli fields with masses around that of the gravitino\cite{Yanagida}, it would push the upper bound on the supersymmetry breaking scale so low that the sparticle spectrum would be unrealistic. It seems that the constraints on the gravitino and potential cosmological moduli fields would severely restrict the parameter space. In our original GMES paper\cite{Gmes}, we propose going in the opposite direction and make the gravitino extremely massive thereby completely decoupling it from the low energy theory. Obviously, one then needs to explain how the gauge hierarchy problem is solved and given the large supersymmetry breaking(Planckian in our case), how one would generate small soft masses. We achieved it by sequestering the hidden sector and visible sector on different branes in a Randall-Sundrum setup\cite{RS1}. The squarks and sleptons would then obtain their masses through visible brane messenger fields that couple to bulk hypermultiplets that carry exponentially suppressed SUSY breaking from the hidden sector. Moreover, GMES also solves the SUSY flavor problem with completely natural order one parameters. The authors of Ref.\cite{YN} have also considered a model with similar low energy degrees of freedom except their gravitino mass is between 100 GeV to 1  TeV which is problematic due to the reasons outlined above.

In this paper, we explore and benchmark a minimal GMES model with the assumption that the observed cold dark matter in the universe consists solely of the neutralino LSP in this model.  For different regions of the parameter space, the low energy spectrum was calculated. Further calculations are then performed to obtain the neutralino dark matter relic density as well as the branching ratios of exotic B-meson decays that could conceivably be enhanced due to the additional supersymmetric particles. Placing the bounds from LEP2 Higgs searches\cite{LEP} and collider searches of supersymmetric partners, the region of parameter space consistent with these bounds were mapped out. The low energy spectrum is very similar to gauge mediation with the only difference being the superheavy gravitino that is completely absent. Because of this, what is normally the next-to-lighest supersymmetric particle (NLSP) in traditional gauge mediation is the LSP in our scenario. In fact, the LSP in our scenario is most often a neutrallino that is dominantly bino-like. The sleptons and squarks exhibit the same mass splittings as what one would expect from gauge-mediated theories whereby colored particles obtain the most significant contributions. The NLSP in our scenario is the stau, which because of its nearly degenerate mass ($\leq 5 \%$) with the neutralino, means that the viable regions of parameter space in our model exhibit the properties of the bino-stau coannihilation region of mSUGRA models. In essence, GMES is a hybrid of both gauge-mediated and mSUGRA models.

The rest of the paper is organized as follows. In Section 2, we review the model presented in Ref.\cite{Gmes}. The calculational procedure for computing the low energy spectrum as well as astrophysical and collider bounds are outlined in Section 3. In Section 4, we present the viable regions of the parameter space and discuss the key features. The differences between the GMES and mSUGRA spectra are also considered. Finally in Section 5, we summarize our findings and conclude.

%%%%%%%%%%%%%%%%%%%%%%%%%%%%%%%%%%%%%%%
\section{Setup}
%%%%%%%%%%%%%%%%%%%%%%%%%%%%%%%%%%%%%%%

In the original GMES scenario\cite{Gmes}, we considered the model
from both sides of the AdS-CFT duality to extract maximal insight
but for most practical purposes, it is sufficient to consider only
the AdS description. The setup of the theory as follows. We have a
five-dimensional Randall-Sundrum\cite{RS1} bulk containing the
supergravity multiplet and hypermultiplets that are needed for
stabilization of the extra-dimension. The hidden sector is on the
UV brane and has Planckian magnitude supersymmetry breaking while
the messenger sector for gauge mediation as well as our visible
sector is localized on the IR brane. The physical separation of
the two sectors would result in the SUSY breaking transmitted to
the IR brane by the massive bulk scalars being exponentially
suppressed. Furthermore, with the bulk warp factor, we can obtain
natural electroweak scale for the soft masses despite starting
with entirely $O(1)$ parameters at the Planck scale.

\begin{figure}
\begin{picture}(200,200)(0,0)
% For branes
\SetScale{2}
\SetColor{Green}
\Line(0,10)(20,20) \Line(20,20)(20,80)
\Line(20,80)(0,90) \Line(0,90)(0,10) \Line(100,10)(80,20)
\Line(80,20)(80,80) \Line(80,80)(100,90) \Line(100,90)(100,10)
% For SUSY mediation
\SetColor{Red} \CArc(100,160)(130,225,265)
% For Labels
\Text(180,20)[]{Messenger and}
\Text(180,10)[]{Visible Sector}
\Text(25,15)[]{Hidden Sector} \Text(25,95)[]{SUSY breaking}
\Text(25,155)[]{UV} \Text(180,155)[]{ IR}
\end{picture}
\caption{
 Schematic diagram illustrating the setup.
}
\end{figure}
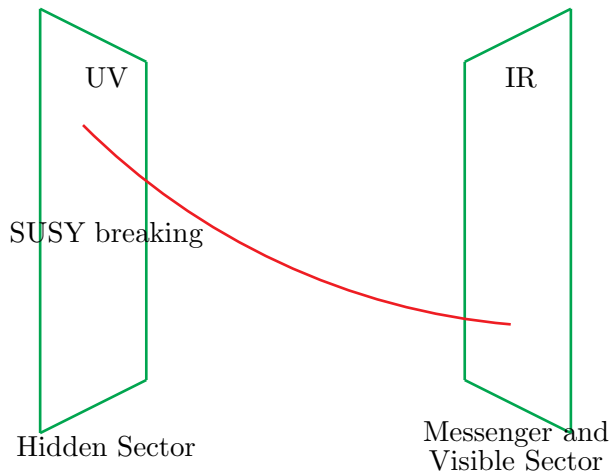

Parametrizing the warp factor by $\omega$, the supersymmetric contributions to the visible sector soft masses coming from the various mediation mechanisms can be calculated. The result from Ref.\cite{Gmes} is
\bea
    m_{\rm soft} \sim \left \{ \begin{array}{ll} \frac{\alpha_{\rm SM}}{4 \pi}
  \Lambda_{IR}  \omega^{\frac{d-5}{2n-3}(n-2)}& gauge  \\
      \Lambda_{IR} \omega^{\frac{d-5}{2n-3}(n-1)} & direct  \\
   \frac{\alpha_{\rm SM}}{4 \pi} \Lambda_{IR}  \omega^{\frac{d-5}{2n-3}n} & anomaly \\
    \frac{1}{16 \pi^2} \Lambda_{IR}  \omega &gravity
    \end{array}\right.
\eea
where $\Lambda_{IR}= M_P \omega$, $d$ is the related to the mass of the bulk hypermultiplets and $n$ is a positive integer parameter in the
IR brane-localized superpotential. The direct contribution arises from corrections to the Kahler potential coming from brane-localized contact terms between the bulk hypermultiplets and Standard Model fields. The anomaly contribution is also  decoupled from the mass of the gravitino in this model\cite{Noscale}. By the gauge contribution, we are referring to the amount of supersymmetry breaking that is transmitted from the UV via the bulk hypermultiplets that couple to the messenger sector which are charged under the Standard Model gauge fields. Specifically, we are considering a brane-localized IR superpotential of the form
\bea
W_{mess} = \Phi \bar{\Phi} H_{IR} \label{Wmess}
\eea
where $H_{IR}$ is the bulk hypermultiplet evaluated at the IR scale and $\Phi$ and $\bar{\Phi}$ are the vector-like pair of messengers
in $\bar{\bf 5} \oplus {\bf 5}$ representation
under the standard model gauge group
SU(5)$_{\rm SM} \supset$ SU(3)$\times$SU(2)$\times$U(1).

In the original paper, we found that
\bea
H_{IR} &\sim & \Lambda_{IR} \omega^{\frac{d-5}{2 n-3}},  \nonumber \\
F_{IR} &\sim  & r \Lambda_{IR}^2 \omega^{\frac{d-5}{2 n-3}(n-1)}.
\eea
where we have introduced a new parameter $r$ to specify
 the ratio $F_{IR}/H_{IR}^2$.
For naturalness, we take $0.1 \leq r \leq 1$.
As discussed in the original paper, there are theoretical constraints
 on model parameters: $n \geq 3$ and $d \leq 7-1/(n-1)$.
The former is required for the messenger fields
 not to break the SM gauge symmetry
 while the latter one is due to theoretical requirement
 that the radius stabilization potential from
 the bulk hypermultiplets be larger than the
 contribution from the Casimir effect.

Assuming a low-energy MSSM content, the gauge mediation
 contribution to the sparticle masses is roughly given by
\bea
  m_{\rm soft} \sim \frac{\alpha_{\rm SM}}{4 \pi}
   \frac{F_{IR}}{\Phi_{IR}}
   \sim 10^{-2} \; r \; \Lambda_{IR} \; \omega^{\frac{d-5}{2 n-3} (n-2)},
\label{GMSBsoftmass}
\eea
with the messenger scale
 $M_{\rm mess} = \Lambda_{IR} \omega^{\frac{d-5}{2 n-3}}$,
 where $\alpha_{\rm SM} \sim 0.01$ stands for the SM gauge coupling constants.

We also have the direct mediation contribution,
\bea
 m_{\rm{direct}} \sim  \frac{F_{IR}}{\Lambda_{IR}}
  \sim r \; \Lambda_{IR} \; \omega^{\frac{d-5}{2 n-3} (n-1)}.
\eea
This is generally flavor-dependent and should be a sub-dominant
 contribution compared to the flavor-blind gauge mediation contribution.
Define the ratio as
\bea
 \epsilon = \frac{m_{\rm{direct}}}{m_{\rm soft}}
         \sim 10^2 \omega^{\frac{d-5}{2 n-3}}.
\eea
Using this and the relation $\Lambda_{IR}  = M_5 \omega$,
 Eq.(\ref{GMSBsoftmass}) leads to the relation
 between $d$ and $\epsilon$,
\bea
  d = 5+ (2 n -3)
  \frac{\log( 10^{-2} \epsilon) }
  {\log \left( \frac{10^{2 (n-1)} m_{\rm soft}}
   {r \epsilon^{n-2} M_5} \right)}.
\eea
We know that the FCNC processes induced by flavor dependent soft terms are strongly
 constrained by experiments, roughly $\epsilon \leq 10^{-2}$
\cite{SUSYFCNC}.
This leads to a strong constraint on the model parameters.
For natural scale of $m_{\rm soft}=$100 GeV$-$1  TeV and
 $ 0.1 \leq r \leq 1$,
 we can find that only $n=3$ (if $n$ is naturally an integer)
 is consistent  with the theoretical constraint $d \leq 7-1/(n-2)$.

In the case $n=3$, we have $m_{\rm soft} \sim 0.01 r M_{\rm mess}$,
 so that 10  TeV $\leq M_{\rm mess} \leq$ 1000  TeV
 for 100 GeV$\leq m_{\rm soft} \leq$ 1  TeV.
Also, we find the composite scale as
 $10^8$ GeV $\leq \Lambda_{IR} \leq$ $1.3 \times 10^{10}$ GeV.
More interestingly, the condition $d \leq 6.5$ leads to
 the lower bound on $\epsilon \geq 0.0016$,
 being an order of magnitude smaller than the current experimental
 bound on SUSY FCNC processes. We expect that future experiments will reveal a sizable FCNC contribution
 originating from flavor-dependent soft masses.
In contrast, conventional gauge mediation models have negligibly
 small FCNC predictions.

%%%%%%%%%%%%%%%%%%%%%%%%%%%%%%%%%%%%%%%
\section{Procedure and Constraints}
%%%%%%%%%%%%%%%%%%%%%%%%%%%%%%%%%%%%%%%

We ran the renormalization group equations using the SuSpect\cite{Suspect} program (version 2.34). This essentially consisted of specifying the variables at the messenger scale, $M_{mess}$, and running it down to obtain the low energy spectrum. In practice, however, this is considerably more involved as the program runs it down with preliminary guesses of $\mu$ and $B \mu$ and back up in energy again to ensure consistent electroweak symmetry breaking. In the process, it checks for an absence of color breaking vacuum and that the Higgs potential is bounded from below as well as the absence of Landau poles. SuSpect iterates the RG running multiple times until it arrives at numerically stable values for electroweak symmetry breaking. With the consistent values, it would then compute the sparticle masses and check whether there are tachyonic colored fields. This, as we shall see, is absolutely crucial as swathes of parameter space have been ruled out simply because of the left-right mixing inducing large off-diagonal terms that ultimately renders the 3rd generation tachyonic. If the above are all satisfied, we then check whether the LSP is a neutralino. This is because we are interested in the LSP in this framework being the thermal relic that constitutes the cold dark matter in the universe. For the rest of the paper, the checks listed in the paragraph would be known as Criteria Set 1.

The resulting low energy spectrum was then fed into the Micromegas\cite{Micromegas} program (version 2.0.1) with the assumption that the low energy effective theory is the MSSM with R-parity  conservation. Micromegas assumes that the decay products of the WIMP are light and in equilibirum, which allows simplification of the Boltzmann equation. The ratio of the WIMP mass to the freeze-out temperature can then be obtained, which in most of our cases is $\approx 25$. Using this value, the relic abundance is calculated taking care to include the possibility of coannihilation with other sparticles. Also included within Micromegas are subroutines that will return the Higgs mass as well as the branching ratios for various exotic B-meson decays. We shall use all these to arrive at the final allowed region of parameter space. Criteria Set 2 is therefore composed of the following bounds.
\begin{itemize}
\item WMAP dark matter relic density\cite{WMAP}:    $\Omega_{\textrm DM} h^2 = 0.0855 - 0.1189$
\item LEP2 Higgs mass bound\cite{LEP}: $m_h \geq 114$ GeV
\item Branching ratio of exotic b decays\cite{PDG, HFAG}: $Br(b \longrightarrow s \gamma)= 2-5 \times 10^{-4}$, $Br(B_s \longrightarrow \mu^+ \mu^-) < 1.5 \times 10^{-7}$.
\end{itemize}

The Heavy Flavor Averaging Group publication\cite{HFAG} actually gives a value of $Br(b \longrightarrow s \gamma)=  (355 \pm 24 ^{+9} _{-10} \pm 3) \times 10^{-6}$
with a reduced $\chi^2= 0.74/4$, where the errors are combined statistical and systematic, systematic due to
the shape function, and the fraction that decays into $d \gamma$. The last two errors are estimated to be the difference of
the average after simultaneously varying the central value of each experimental result by $\pm 1 \sigma$. Given that this is not the $2\sigma$ value and also the inherent theoretical uncertainties from QCD corrections that accompany the Micromegas calculation, we have decided, as is conventionally done, to choose the limit in Criteria Set 2. It should be noted that the WMAP dark matter density and LEP2 Higgs mass bound in combination would restrict parameter space to a region with a branching ratio to $3-4 \times 10^{-4}$.

We will not be be considering in detail the direct detection bounds coming from CDMS2\cite{CDMS} as our neutralino LSP is dominantly bino-like. THe CDMS2 bound on spin-independent elastic-scattering cross section is in the $10^{-6}$ picobarn range while in our viable regions, it is $10^{-9}-10^{-8}$ picobarns.

%%%%%%%%%%%%%%%%%%%%%%%%%%%%%%%%%%%%%%%
\section{Parameter Space of GMES}
%%%%%%%%%%%%%%%%%%%%%%%%%%%%%%%%%%%%%%%

Using SuSpect and Micromegas, we investigated the parameter space of GMES for the minimal gauge-mediated sector (i.e. a messenger superpotential in the form of Eq.(\ref{Wmess})) consisting of either one or two complete SU(5) messenger multiplets and \tanb of 10, 30 and 50. While our parameter space is considerably narrower than Ref.\cite{Nomura} because of both phenomenological and theoretical constraints, it is nevertheless consistent with their findings in the common region. In this section we shall also compare and comment of the generic spectrum arising from GMES with that of CMSSM(the phenomenological model derived from mSUGRA) which in its simplest incarnation also does not have a gravitino in its low energy spectrum.

\begin{figure}[t]
\includegraphics[scale=0.5]{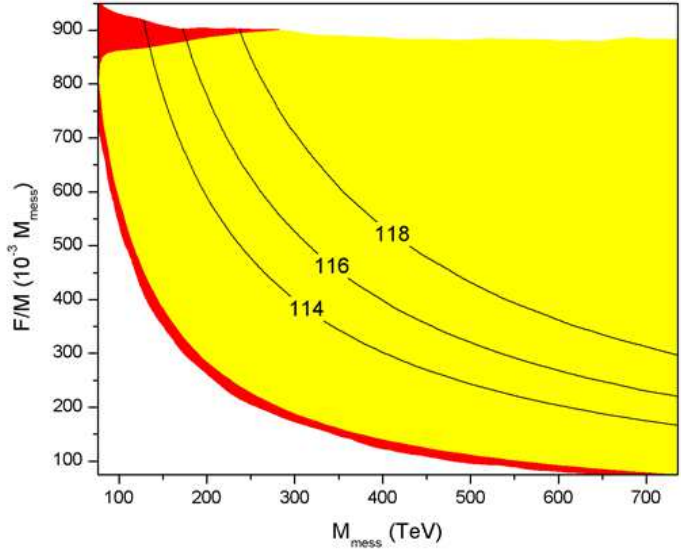}
\caption{Parameter space plot for one messenger multiplet of $5 \oplus \bar{5}$,  $\mu>0$, \tanb = 10.   The colored regions are that which are consistent with Criterion Set 1; i.e. consistent RG running, neutralino LSP and viable EWSB. The region in red is that which satisfies the dark matter relic abundance as well. Finally, the contours are that of the Higgs mass where the number gives the mass in GeV.  }\label{t10}
\end{figure}

\subsection{One messenger multiplet of $5 \oplus \bar{5}$,  $\mu>0$ and $\tanb = 10$}
Among the cases we considered, this has the largest area of parameter space (the colored regions of Figure 2) consistent with Criteria Set 1 above; RG running, EWSB breaking, neutralino LSP. The parameter space above the viable region is excluded due to the fact that the LSP in these scenarios is the stau rather than the neutralino. As in conventional gauge mediation, the left-right mixing is enhanced by the tau Yukawa coupling and the mass of lightest one can be pushed below that of the lightest neutralino. The bottom left corner of the uncolored parameter space is excluded because of the tachyonic third generation sfermions coming from large off-diagonal terms in the left-right mixing. Closer but still to the left of the viable region, we can obtain a EWSB-consistent low energy spectrum, but the resultant sparticle spectrum is too light and is excluded by collider bounds. This can be easily seen from estimates of the sparticle masses from $\frac{\alpha_{SM} F}{4 \pi M}$.

When we apply the dark matter relic abundance constraint, we find a narrow sliver of parameter space remains (the red region of Figure 2). In fact, if we look at the left edge of this red region, we find that the relic density is higher than the WMAP lower bound, i.e. $> 0.0855$. So it is the collider bounds that cut off the parameter space on the left side. Moreover, the dark matter relic abundance increases are we move to higher values of $M_{mess}$ while keeping the $F/M_{mess} ^2$ fixed. In the red region that extends down in a crescent-like form, the neutralino-neutralino annihilation processes are at work and with increasing LSP mass, the cross-section drops and hence the relic density increases. The nearly horizontal part at the top of the red region is rather more interesting. This is where the bino-stau coannihilation dominates. Actually, as we will later see in a representative point, the also near degeneracy of selectron and smuon with the bino-like neutralino means that these are also effective channels.

And the LEP2 constraints\cite{LEP} on the Higgs mass also rules out most of the red region forcing us into this tiny region of the parameter space. We can see the the near degeneracy of neutralino (with eigenvector \{0.991, -0.028, 0.117, -0.058\} in the basis of \{bino, wino, lighter higgsino, heavier higgsino\}) and stau masses (a mass splitting of 8 GeV) in the spectrum of a GMES point in Table 1 ($M_{mess} =$ 140  TeV, $\fom  =$ 123  TeV). Note also the 10 GeV mass splitting between the neutralino and selectron and smuon. For this point in the parameter space (which gives us $\Omega_{\textrm DM} h^2 = 0.114$), neutralino-stau coannihilation to tau and photon dominates followed by neutralino-selectron to electron and photon  and neutralino-smuon to muon and photon on equal footing. This is probably not the most representative point in the region as we tried to keep the gluinos light. Most of the points in the viable region have neutralino-tau mass splittings well less than 8 GeV so the coannihilation process is even more effective.

%%%%%%%%%%%%%%%%%%%%%%%%%%%%%%%%
% Table I
%%%%%%%%%%%%%%%%%%%%%%%%%%%%%%%%
\begin{table}[t]
\begin{ruledtabular}
\begin{tabular}{c|ccc}
Mediation      & GMES     & CMSSM A        & CMSSM B  \\
                 & $M_{mess} =$ 140  TeV      & $m_0 =  $ 978   & $m_0 =  $ 272 \\
                & $\fom  =$ 123  TeV     & $m_{1/2} =  $ 472     & $m_{1/2} =  $ 496\\
\hline
$m_h$                    & 114       & 116   & 115  \\
$m_H$                    & 640    & 1170          & 751 \\
$m_A$                    & 640       & 1170     &  751 \\
$m_{H^{\pm}}$       & 645      & 1173          &   755\\
\hline
$m_{\tilde{\chi}^{\pm}_{1,2}}$ & 368, 521        & 371, 627       &  384, 648\\
$m_{\tilde{\chi}^0}$     & 199, 369, 491, 521     & 196, 371, 613, 627      & 204, 385, 634, 648 \\
$m_{\tilde{g}}$   & $\underline{1144} $    & $\underline{1144} $    & $\underline{1144} $   \\
\hline
$m_{\tilde{t}_{1,2}}$    & 1211,1338      &    953, 1219       &  812, 1024\\
$m_{{\tilde{u},\tilde{c}}_{1,2}}$  & 1316, $\underline{1370}$    & 1349, $\underline{1370} $   & 1035, 1070  \\
\hline
$m_{\tilde{b}_{1,2}}$    & 1307, 1329   & 1199, 1336   & 984, 1027\\
$m_{{\tilde{d},\tilde{s}}_{1,2}}$    &  1311, 1372	& 1347, 1372    & 1032, 1073 \\
\hline
$m_{\tilde{\tau}_{1,2}}$  & 207, 428  & 983, 1019 & 324, 429 \\
$m_{{\tilde{e},\tilde{\mu}}_{1,2}}$   & 209, $\underline{428}$     & 992, 1023     & 330, $\underline{428}$  \\
\end{tabular}
\end{ruledtabular}
\caption{
Sparticle and Higgs boson mass spectra (in units of GeV)
 in the GMES and the CMSSM, with $\tanb =10$.
In each CMSSM, input universal gaugino mass ($M_{1/2}$)
 at the GUT scale ($M_{GUT}=2 \times 10^{16}$ GeV)
 has been taken so as to give the same gluino mass
 as in the GMES.
Universal soft mass squared in each CMSSM  was taken
 so as to give the same squark and slepton masses
 as in the GMES. In the CMSSM cases, we took $A=0$.
}
\label{table1}
\end{table}

%%%%%%%%%%%%%%%%%%%%%%%%%%%%

Let us now check a point in the crescent-like region, say $M_{mess}$=200 TeV and \fom=54 TeV. We obtained $\Omega_{\textrm DM} h^2 = 0.110$ and the neutralino and stau masses are 69 and 98 GeV respectively. Neutralino-neutralino annihilation into tau-leptons, muons and electrons accounting for nearly all the annihilation. So this is indeed the usual annihilation process but it is unfortunately excluded by LEP2 as can be seen from the Higgs mass contours.

As for the exotic B-meson decay constraints, the points that satisfy the dark matter relic density and LEP2 Higgs bound automatically satisfy these as well. In the $Br(b \longrightarrow s \gamma)$  case, the values fall between $3.7-3.9 \times 10^{-4}$ while for $Br(B_s \longrightarrow \mu^+ \mu^-)$ we get a value of $\approx 3.1 \times 10^{-9}$.

To do a comparison between CMSSM and GMES,  consider a phenomenologically viable point in the red region, $M_{mess} =$ 140  TeV, $\fom  =$ 123  TeV. Using the same \tanb for both, let us vary two parameters in the CMSSM case; $m_0$  and $m_{1/2}$ which controls the soft masses of the sfermions and gauginos respectively. This would then allow us to match two masses on both sides of the comparison. Since we are varying $m_{1/2}$, we used the same gluino mass as a starting point. Additionally, we will also match a slepton as well as a squark. Our results are presented in Table 1. We see that the slepton masses in CMSSM A are considerably higher which is to be expected as the GMES sleptons have masses proportional to their gauge couplings. The CMSSM A squarks on the other hand have comparable masses to the GMES case. The other point to note is that the other Higgs masses are considerably higher in the CMSSM A case despite having approximately the same mass for the lightest Higgs. This is due to our requirement that we have the correct electroweak symmetry breaking and therefore the mass of the lightest Higgs is fine-tuned. The other Higgses on the other hand would receive masses proportional to the ``generic'' scale of SUSY breaking. By that, we mean that since the gluino masses are matched in both scenarios, GMES has a lower ``generic'' scale of SUSY breaking (i.e. $F/M_{mess}$ is less than $m_{1/2}$) as there is an enhancement coming from the gauge coupling. As a result, the SUSY breaking seen by non-colored fields in the GMES scenario are much lighter. Finally, we consider the case where the CMSSM spectra has the same selectron mass. We see a drop in the scales of the masses as compared to the previous scenarios. This is simply due to the fact that CMSSM mediates roughly the same masses to all the fields while GMES apportions the masses according to their coupling to gauge fields. As can be inferred from our above discussion, once we match the gluino and selectron, the CMSSM would have much lower squark masses than GMES.

\begin{figure}[t]
\includegraphics[scale=0.5]{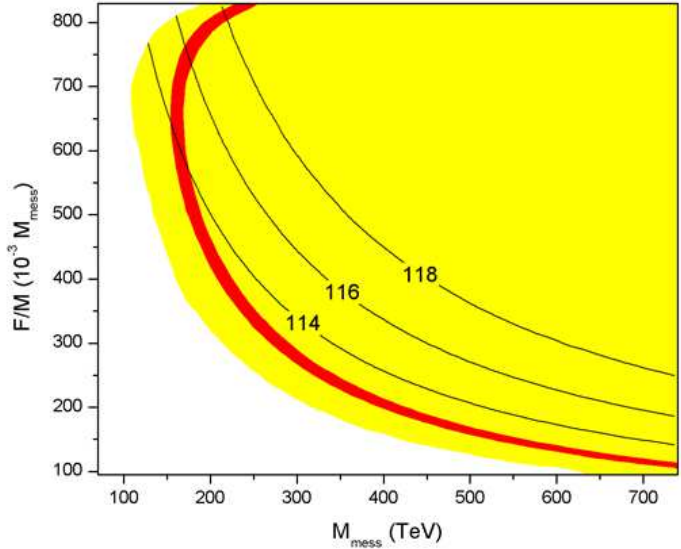}
\caption{
Parameter space plot for one messenger multiplet of $5 \oplus \bar{5}$, $\mu>0$, \tanb = 30.   The colored regions are that which are consistent with Criterion Set 1. The region in red is that which satisfies the dark matter relic abundance as well. Finally, the contours are that of the Higgs mass where the number gives the mass in GeV.}
\label{t30}
\end{figure}

\subsection{One messenger multiplet of $5 \oplus \bar{5}$, $\mu>0$ and $\tanb = 30$}

This case has a smaller parameter space (colored region of Figure 3) consistent with RG, EWSB and neutralino LSP as compared to the previous scenario. As before, the upper \fom region is excluded  due to the fact that we get a stau LSP there. While the points on the far left of the excluded parameter space (i.e. small $M_{mess}$ and small \fom) simply do not give sparticles sufficiently heavy masses, the points slightly just to the left and bottom of the viable RG region are in fact excluded because they give a stau LSP. In this scenario, there are also quite a significant number of points in the left corner of the parameter space ruled out due to a tachyonic third generation of sfermions coming from large off-diagonal terms in the left-right mixing.

Once we apply the relic density constraint, we get a narrow region of parameter space remaining (the red region). Now the yellow region to the left of the red sliver is theoretically still possible provided we have another source of dark matter but we will not explore this possibility further in the present paper. Another point to note is that as we move down the red sliver, ignoring the LEP2 Higgs bound, the neutralino-stau coannihilation becomes less important as we will show explicitly with a representative point further in this subsection.

%%%%%%%%%%%%%%%%%%%%%%%%%%%%%%%%
% Table 2
%%%%%%%%%%%%%%%%%%%%%%%%%%%%%%%%
\begin{table}[t]
\begin{ruledtabular}
\begin{tabular}{c|ccc}
Mediation      & GMES     & CMSSM A        & CMSSM B  \\
                 & $M_{mess} =$ 165  TeV      & $m_0 =  $ 831   & $m_0 =  $ 264 \\
                & $\fom  =$ 104  TeV     & $m_{1/2} =  $ 362     & $m_{1/2} =  $ 382\\
\hline
$m_h$                    & 114       & 115   & 114  \\
$m_H$                    & 470    & 780          & 497 \\
$m_A$                    & 470      & 780     & 497 \\
$m_{H^{\pm}}$       & 477     & 784   &   504\\
\hline
$m_{\tilde{\chi}^{\pm}_{1,2}}$ & 284, 454        & 279, 484       &  292, 510\\
$m_{\tilde{\chi}^0}$     & 151, 284, 434, 454     & 149, 281, 470, 486     & 155, 292, 495, 509 \\
$m_{\tilde{g}}$   & $\underline{901} $    & $\underline{901} $    & $\underline{901} $   \\
\hline
$m_{\tilde{t}_{1,2}}$    & 1025, 1127      &    768, 960       & 641, 814 \\
$m_{{\tilde{u},\tilde{c}}_{1,2}}$  & 1116, $\underline{1118}$    & 1104, $\underline{1118} $   & 856, 830 \\
\hline
$m_{\tilde{b}_{1,2}}$    & 1084, 1124   & 930, 1023  & 749, 804\\
$m_{{\tilde{d},\tilde{s}}_{1,2}}$    &  1115, 1169	& 1103, 1121   & 828, 860 \\
\hline
$m_{\tilde{\tau}_{1,2}}$  & 159, 373 & 766, 832  & 251, 372\\
$m_{{\tilde{e},\tilde{\mu}}_{1,2}}$   & 182, $\underline{369}$  & 841, 862       & 302 , $\underline{369}$ \\
\end{tabular}
\end{ruledtabular}
\caption{
Sparticle and Higgs boson mass spectra (in units of GeV)
 in the GMES and the CMSSM, with $\tanb=30$.
In each CMSSM, input universal gaugino mass ($M_{1/2}$)
 at the GUT scale ($M_{GUT}=2 \times 10^{16}$ GeV)
 has been taken so as to give the same gluino mass
 as in the GMES. Universal soft mass squared in each CMSSM  was taken
 so as to give the same squark and slepton masses
 as in the GMES. In the CMSSM cases, we took $A=0$.}
\label{table2}
\end{table}
%%%%%%%%%%%%%%%%%%%%%%%%%%%%

The LEP2 bounds are once again effective in narrowing down the region of parameter space. For the GMES point in Table 2 ($M_{mess} =$ 164  TeV, $\fom  =$ 104  TeV), so chosen because we wanted to keep the gluinos light while still satisfying the LEP2 bounds, we find that the neutralino dark matter density is $\Omega_{\textrm DM} h^2 = 0.109 $ and the neutralino has an eigenvector of \{0.991, -0.027, 0.124, -0.047\} in the basis of \{bino, wino, lighter higgsino, heavier higgsino\}. Looking at the spectrum, we can already deduce from the 8 GeV splitting in the neutralino and stau masses that we are in the neutralino coannihilation region. Explicit calculation gives the dominant contribution as neutralino-stau coannihilation to photon and stau.

For purposes of consideration, let us take a point further down on red sliver which violates the LEP2 Higgs bound, say $M_{mess}$=540 TeV and \fom=81 TeV, and we find that the LSP is also dominantly bino-like. We obtained $\Omega_{\textrm DM} h^2 = 0.0993$ and the neutralino and stau masses are 108 and 118 GeV respectively. The annihilation process of neutralino-stau into tau and photons is significant but neutralino-neutralino to taus, bottom quarks, muons and electrons accounts for the largest fraction respectively. This is similar to the A-annihilation funnel region of mSUGRA where $\chi_0 +\chi_0 \longrightarrow A \longrightarrow b \bar{b}, \tau  \bar{\tau},$ etc. This would be even more pronounced when we go to higher \tanb as we shall see.

As in the previous case, the region that satisfies the relic density and LEP2 constraints automatically satisfies the $Br(b \longrightarrow s \gamma)$  and $Br(B_s \longrightarrow \mu^+ \mu^-)$ bounds with values of $3.7-3.8 \times 10^{-4}$ and $3.0-3.3 \times 10^{-9}$ respectively.

Table 2 gives the spectrum of a representative GMES point ($M_{mess} =$ 164  TeV, $\fom  =$ 104  TeV) as well as that of CMSSM (matched to specific masses of GMES) for $\tanb = 30$. We are also requiring that gluino masses be the same within all the columns. As is expected from lowering $\fom$ (as compared to Table 1), the sparticle masses in this case are lower. We observed the same trends as in Table 1, as in sleptons-squarks splitting in GMES is far greater than the CMSSM cases because in GMES, the splitting arises due to quantum numbers of the fields while in CMSSM, the splittings are generated from RG running. Once again, we see the near degeneracy of the lightest neutralino and stau and this is at a point far from the completely degenerate edge of the parameter space. So most of the points would likely have a neutralino-stau splitting less than 8 GeV.

\begin{figure}[t]
\includegraphics[scale=0.5]{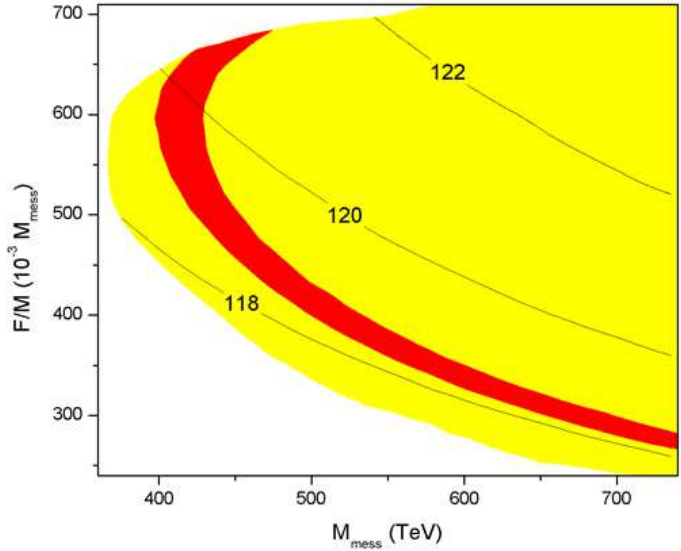}
\caption{
Parameter space plot for one messenger multiplet of $5 \oplus \bar{5}$, $\mu > 0$, \tanb = 50.  The colored regions are that which are consistent with Criterion Set 1. The region in red is that which satisfies the dark matter relic abundance as well. Finally, the contours are that of the Higgs mass where the number gives the mass in GeV.}
\label{t50}
\end{figure}

\subsection{One messenger  multiplet of $5 \oplus \bar{5}$,  $\mu>0$ and $\tanb = 50$}

This scenario has an even smaller region that satisfies Criteria Set 1. The exclusion is mainly due to the stau LSP or the tachyonic 3rd generation of sfermions as a result of left-right mixing. The dark matter relic abundance once again restricts the parameter space to a sliver but the LEP2 bounds on the Higgs mass does restrict the parameter space in any way. So let us consider the GMES point in Table 3 that somewhat minimizes the mass of the gluino, $M_{mess} =$ 560  TeV, $\fom  =$ 205  TeV. The resultant neutralino LSP is composed of \{0.998,-0.007,0.066,-0.025\} in the basis of \{bino, wino, lighter higgsino, heavier higgsino\}. We obtained $\Omega_{\textrm DM} h^2 = 0.106$ and the neutralino and stau mass splitting of 9 GeV. Despite the small splitting, we find that the dominant annihilation channels are neutralino-neutralino to bottom quarks followed by neutralino-stau to photon and tau. This is a combination of coannihilation together with nearly A-resonance annihilation where $\chi_0 +\chi_0 \longrightarrow A \longrightarrow b \bar{b}$ which is enhanced at high \tanb.

For purposes of consideration, let us choose another point further up the sliver, say $M_{mess}$=430 TeV and \fom=280 TeV, to see if the neutralino-stau coannihilation ever dominates.
The LSP is overwhelmingly bino-like and we obtained $\Omega_{\textrm DM} h^2 = 0.107$. Indeed, the dominant annihilation channel is neutralino-stau to photon and tau. 

As before, the branching ratios of $b \longrightarrow s \gamma$  and $B_s \longrightarrow \mu^+ \mu^-$ are automatically satisfied with values of $3.8-3.9 \times 10^{-4}$ and $3.9-6.9 \times 10^{-9}$ respectively.

%%%%%%%%%%%%%%%%%%%%%%%%%%%%%%%%
% Table 3
%%%%%%%%%%%%%%%%%%%%%%%%%%%%%%%%
\begin{table}[t]
\begin{ruledtabular}
\begin{tabular}{c|ccc}
Mediation      & GMES     & CMSSM A        & CMSSM B  \\
                 & $M_{mess} =$ 560  TeV      & $m_0 =  $ 1715   & $m_0 =  $ 549 \\
                & $\fom  =$ 205  TeV     & $m_{1/2} =  $ 663.5     & $m_{1/2} =  $ 702.5\\
\hline
$m_h$                    & 119       & 119   & 118  \\
$m_H$                    & 472    & 210         & 462 \\
$m_A$                    & 472       & 210    &  462\\
$m_{H^{\pm}}$       & 480      & 229          &   470\\
\hline
$m_{\tilde{\chi}^{\pm}_{1,2}}$ &  549, 806       & 535, 740       &  561, 836\\
$m_{\tilde{\chi}^0}$     & 287, 549, 794, 806    & 284, 535, 724, 740      & 297, 561, 824, 835 \\
$m_{\tilde{g}}$   & $\underline{1585} $    & $\underline{1585}  $    & $\underline{1585}  $   \\
\hline
$m_{\tilde{t}_{1,2}}$    & 1859, 2004     &    1456, 1625       &  1159, 1334\\
$m_{{\tilde{u},\tilde{c}}_{1,2}}$  & 2048, $\underline{2148}$    & 2122, $\underline{2148} $   & 1476, 1525  \\
\hline
$m_{\tilde{b}_{1,2}}$    & 1902, 2014   & 1598, 1664  & 1246, 1334 \\
$m_{{\tilde{d},\tilde{s}}_{1,2}}$    &  2038, 2149	& 2120, 2149    & 1471, 1527 \\
\hline
$m_{\tilde{\tau}_{1,2}}$  & 296, 717  & 1231, 1543 & 399, 668 \\
$m_{{\tilde{e},\tilde{\mu}}_{1,2}}$   & 358, $\underline{718}$    & 1730, 1762      & 607, $\underline{718}$  \\
\hline
\end{tabular}
\end{ruledtabular}
\caption{
Sparticle and Higgs boson mass spectra (in units of GeV)
 in the GMES and the CMSSM, with $\tanb=50$.
In each CMSSM, input universal gaugino mass ($M_{1/2}$)
 at the GUT scale ($M_{GUT}=2 \times 10^{16}$ GeV)
 has been taken so as to give the same gluino mass
 as in the GMES. Universal soft mass squared in each CMSSM  was taken
 so as to give the same squark and slepton masses
 as in the GMES. In the CMSSM cases, we took $A=0$.}
\label{table3}
\end{table}
%%%%%%%%%%%%%%%%%%%%%%%%%%%%

Table 3 gives the spectrum of a representative GMES point ($M_{mess} =$ 560  TeV, $\fom  =$ 205  TeV) as well as that of CMSSM (matched to specific masses of GMES) for $\tanb = 50$. We required that gluino masses be the same within all the columns. As is expected from raising $\fom$ (as compared to Table 1 and 2), the sparticle masses in this case are higher. We observed the same trends as in Table 1 and 2, as in sleptons-squarks splitting in GMES is far greater than the CMSSM cases because in GMES, the splitting arises due to gauge coupling while in CMSSM, the splittings are generated from RG running. The near degeneracy of the lightest neutralino and stau pointedly indicates a strong coannihilation channel. What is perhaps less obvious is the fact that the GMES point in question is in what is known to mSUGRA afficionadoes as the A-annihilation funnel. The astute reader might also notice the rather small heavy Higgs masses for the CMSSM A case. The tree-level mass relations no longer even approximately hold in this region because of large corrections from the large $\tan \beta$ in this particular region of the parameter space.

\subsection{Two messenger multiplets of $5 \oplus \bar{5}$, $\mu>0$ and $\tanb = 10, 30, 50$}
The entire parameter space for $\tanb=30 , 50$ is completely excluded due to either the stau LSP or the tachyonic 3rd generation of sfermions coming from left-right mixing. While a very small region of $\tanb=10$ passed Criteria Set 1, the dark matter relic density is however too small with $\Omega_{\textrm DM} h^2 = 0.01 - 0.04$.

%%%%%%%%%%%%%%%%%%%%%%%%%%%%%%%%%%%%%%%
\section{Conclusion}
%%%%%%%%%%%%%%%%%%%%%%%%%%%%%%%%%%%%%%%%
We have explored the allowed regions of the parameter space of the ``Gauge Mediation from Emergent Supersymmetry" scenario. Assuming that the observed dark matter consists solely of a neutralino lightest supersymmetric particle, we calculated the low energy spectra of the parameter space of GMES and find that our neutralino is dominantly bino-like. Applying the relic density constraint  and LEP2 Higgs mass bound, we narrowed down the parameter space to a very small region in all the cases that we considered. Other constraints such as exotic B-meson decay processes and supersymmetric partner searches in colliders were automatically satisfied in the above region. We found that the dominant annihilation channel is that of the neutralino-stau coannihilation with significant contributions from neutralino-selectron and neutralino-smuon chanels, which is reminiscent of mSUGRA scenarios. As we go to higher $\tan \beta$, the channel where neutralino-neutralino decay into the CP-odd Higgs which then decays into bottom quarks, taus, etc becomes more pronounced. In mSUGRA, this is identified as the A-annihilation funnel region. Yet fundamentally, the low energy spectrum is exactly identical to gauge mediation except that the gravitino in our case is superheavy and therefore decouples from the low energy theory. So GMES has a very intriguing set of hybrid properties that challenges conventional understanding of gauge-mediated and minimal supergravity models and it would be very interesting to further constrain this model with future data from the LHC, the PLANCK satelite, high precision exotic B-meson decay experiments and the next generation of direct dark matter searches.

\begin{acknowledgments}
 S.P.N. was supported by the National Science and Engineering Research Council of Canada while N.O. was supported in part by the Grant-in-Aid for Scientific Research from the Ministry of Education, Science and Culture of Japan. The authors would also like to thank Andrew Blechman, Santiago De Lope, Hock-Seng Goh, Hideo Itoh, Shigeki Matsumoto, Mihoko Nojiri, Erich Poppitz, Yanwen Shang, Michisa Takeuchi and Scott Watson for useful discussions.
\end{acknowledgments}

%%%%%%%%%%%%%%%%%%%%%%%%%%%%%%%%%%%%%%%%%%%%%%%%%%%

%
\end{document}